Multicomponent MP4 and the inclusion of triple excitations in multicomponent many-body methods

O. Jonathan Fajen, Kurt R. Brorsen*

Department of Chemistry, University of Missouri, Columbia, Missouri 65203, USA

*brorsenk@missouri.edu

**Abstract**

This study implements the full multicomponent third-order (MP3) and fourth-order (MP4) many-body perturbation theory methods for the first time. Previous multicomponent studies have only implemented a subset of the full contributions and the present implementation is the first multicomponent many-body method to include any connected triples contribution to the electron-proton correlation energy. The multicomponent MP3 method is shown to be comparable in accuracy to the multicomponent coupled-cluster doubles method for the calculation of proton affinities, while the multicomponent MP4 method is of similar accuracy as the multicomponent coupled-cluster singles and doubles (CCSD) method. From the results in this study, it is hypothesized that the relative accuracy of multicomponent methods is more similar to their single-component counterparts than previously. It is demonstrated that for multicomponent MP4, the fourth-order triple-excitation contributions can be split into electron-electron and electron-proton contributions and the electron-electron contributions ignored with very little loss of accuracy of protonic properties.

## I. Introduction

For a class of methods that was first introduced in the late 1960s,[1-3] the rapid progress[4] in multicomponent method development since the introduction[5-7] of the epc17 electron-proton correlation functionals for multicomponent density functional theory (DFT) has been striking, as a variety of additional multicomponent methods such as second-order Møller-Plesset perturbation theory (MP2),[8-9] coupled cluster (CC) theory with single and double excitations (CCSD),[8, 10-16] and configuration interaction (CI)[11, 17-18] are now able to compute protonic properties accurately.[16, 18-20] While the recent proliferation of multicomponent methods has been an absolute success, the multicomponent framework where the reference wave function is written as a product of electronic and protonic Slater determinants is still not well understood. It has been demonstrated that one-shot corrections to multicomponent Hartree-Fock (HF) do not give accurate protonic properties in the case of multicomponent DFT[21-22] and MP2.[16] The bad performance of multicomponent MP2 has been attributed to the poor quality of the multicomponent HF orbitals. Due to the attractive interaction between electrons and nuclei, the protonic density is severely overlocalized in multicomponent HF relative to numerically exact grid-based calculations. This in turn affects most other protonic properties and it has been assumed that multicomponent HF is a qualitatively incorrect zeroth-order reference wave function.

In single-component electronic structure theory for systems in which the zeroth-order wave function is qualitatively correct, many-body theory is one of the most successful ways of including the electron-electron correlation energy. In the development of single-component many-body theory, the single-component wave function and energy was commonly analyzed in terms of single-component many-body perturbation theory.[24-25] Using such analysis, it can be shown that single-component CCD is correct through first order in the wave function and third order in the energy and that single-component CCSD is correct through first order in the wave function and fourth order in the energy except for the neglect of fourth-order triple-excitation energy contributions. In this respect, single-component CCSD is similar to single-component fourth-order many-body perturbation theory with single, doubles, and quadruples (MP4SDQ),[26] though single-component CCSD also contains many higher-order terms that are not present in single-component MP4SDQ. Historically, the decision to neglect the triples contributions to the fourth-order energy in single-component MP4SDQ was made because the computational scaling of the fourth-order singles, doubles, and quadruples terms are all $N^6$ or less with respect to system size, while the triples-excitation term scales as $N^7$. However, after the introduction of the MP4SDQ method, it was soon shown that the fourth-order triples contribution was not small.[27] This realization was part of the impetus for the rapid development of single-component CC methods that include triple excitations,[28-33] which has also led to the many approximate ways of incorporating the contribution from connected triples to correct single-component CCSD including the "gold-standard" method of *ab initio* quantum chemistry, CCSD(T).[34]

Multicomponent many-body methods were first implemented in 2003 by Nakai and Sodeyama.[8] Their study implemented the multicomponent MP2, MP3, MP4, and CCSD methods and tested the methods on a series of diatomic systems. Also important to the present study was their introduction of diagrammatic equations for multicomponent many-body methods using an approach based on Goldstone diagrams. However, a careful examination of their MP3 derivation for the electron-proton correlation energy shows that diagrams are missing. Additionally, due to the number of diagrams that arise in the fourth-order energy, they only included contributions to the electron-proton correlation energy from select ladder- and ring-diagrams for a total of six diagrams, which is substantially less than the complete set of 192 fourth-order electron-proton diagrams. Owing to this small number of diagrams, it is our opinion that the multicomponent MP4 method has not yet been truly implemented.

In principle, an analysis of the multicomponent many-body wave function can be performed completely analogous to the single-component case. While no explicit analysis in terms of the orders of the energy and wave function contributions was ever performed in the original or in subsequent multicomponent many-body studies, it is simple to show that multicomponent CCD is correct through third order in the multicomponent many-body perturbation theory energy and that multicomponent CCSD is correct through fourth order in the energy except for the neglect of fourth-order triple-excitation energy contributions. Both orders are identical to their single-component many-body theory counterparts. However, because the previous implementation of the multicomponent MP4 method only included a limited number of diagrams, the accuracy of the multicomponent MP4SDQ method has never been assessed nor has the importance of the fourth-order triple excitation energy terms ever been investigated.

Given the qualitative failure of multicomponent HF, which is used as a zeroth-order reference for multicomponent many-body perturbation theory, it would seem unlikely that multicomponent MP4 would give accurate results. As we show in this study, this is not the case and a version of multicomponent MP4SDQ is essentially as accurate as multicomponent CCSD. Furthermore, we show that triple-excitation contributions to the fourth-order electron-proton correlation energy significantly decrease the proton affinity error for all calculations in this study. These results strongly suggest that the inclusion of connected triple excitations in multicomponent CC methods is likely to significantly increase the accuracy of such methods in a manner similar to that of single-component CC methods and that it may be possible to only include connected triple contributions involving mixed electron-proton excitations, which would greatly increase computational efficiency.

**II. Theory**

Though they have been previously been derived many times in the literature, we start by introducing the multicomponent MP2 equations to illustrate our multicomponent diagrammatic notation as it differs slightly from the previous notation of Nakai and Sodeyama[8] due to our focus on Hugenholtz diagrams[35] instead of Goldstone diagrams.[36-37] The choice of Hugenholtz diagrams is made because it is simpler to ensure all diagrams have been enumerated when deriving the fourth-order many-body perturbation theory energy.[25] As the number of diagrams in a multicomponent framework is much greater than in a single-component framework, this simplicity is essential in correctly deriving the method. Both our diagrammatic notation and the previous multicomponent diagrammatic notation largely mirror that of single-component many-body theory[25] with only small changes to account for the multiple types of quantum particles. In the following, we assume a single quantum proton is treated quantum mechanically, as has typically been the case for previous multicomponent method studies. In principle, the multiple proton case is straightforward to derive using the rules introduced in this study. Because only a single-proton is treated quantum mechanically, after the introduction of the rules for converting the well-known single-component Hugenholtz diagrams for electron-electron correlation into multicomponent Hugenholtz diagrams, we only focus on diagrams that include electron-proton correlation as no diagrams for proton-proton correlation occur and the diagrams for electron-electron correlation are identical to those in single-component many-body perturbation theory.

We adopt the convention where the indices *i,j,k,…* indicate occupied electronic orbitals, *a,b,c,…* indicate virtual electronic orbitals, and *p,q,r,…* indicate any type of electronic orbital. Protonic orbitals are defined analogously, but with capital letters. We denote a multicomponent configuration as $|\Psi\rangle = |\Psi^{\text{elec}}\rangle \otimes |\Psi^{\text{prot}}\rangle$, where $|\Psi^{\text{elec}}\rangle$ is an electronic Slater determinant and $|\Psi^{\text{prot}}\rangle$ is a protonic Slater determinant. The multicomponent cluster operators are defined as $T_n^{\rho_1 \dots \rho_n} = \sum_{n_i} t_{n_i} \tau_{n_i}$, where $\tau_{n_i}$ is an excitation operator of order *n* and $n_i$ is a label of the holes and particles involved in the excitation. $t_{n_i}$ is the corresponding cluster amplitude. The $\rho_1 \dots \rho_n$ superscript indicates the particle type of the order *n* excitations; i.e., $T_2^{\text{ep}}$ is the double-excitation cluster operator for one-electron, one-proton excitations and $T_3^{\text{eep}}$ is the triple-excitation operator for two-electron, one-proton excitations.

In single-component MP2, there is one directed Hugenholtz diagram and one corresponding antisymmetrized Goldstone diagram (Figure 1), which is interpreted algebraically as

$$E_{\text{MP2}}^{\text{ee}} = \frac{1}{4} \sum_{ijab} \frac{|\langle ij||ab\rangle|^2}{\epsilon_i + \epsilon_j - \epsilon_a - \epsilon_b}, \qquad (1)$$

where the superscript "ee" indicates that single-component MP2 only includes electron-electron correlation.

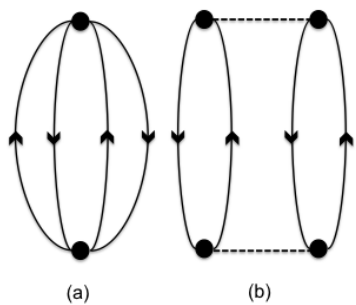

**Figure 1:** (a) Hugenholtz and (b) corresponding antisymmetrized Goldstone diagrams for single-component MP2.

To convert these diagrams to describe electron-proton correlation, we start with all possible directed electron-electron correlation Hugenholtz diagrams and change electronic hole or particle lines to protonic hole or particle lines with the restriction that the number of electronic hole lines is equal to the number of electronic particle lines at each interaction vertex. The same must be true for protonic hole and particle lines. Additionally, for the case of a single proton, there can be no interaction vertex with two protonic hole and two protonic particle lines as this would be interpreted as a proton-proton interaction. We adopt the convention where electronic hole and particle lines are indicated with solid lines and protonic hole and particle lines are indicated with dashed lines. Using these rules, the MP2 electron-proton correlation energy consists of a single Hugenholtz diagram (Figure 2a).

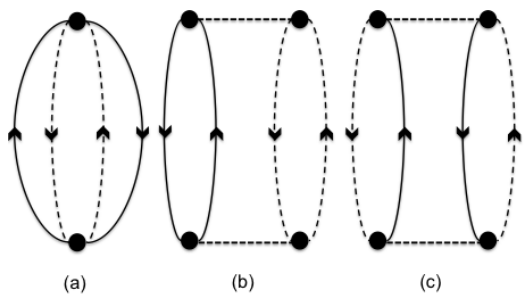

**Figure 2:** (a) Hugenholtz and (b)-(c) two possible expansions of the Hugenholtz diagram in terms of antisymmetrized Goldstone diagrams for the multicomponent MP2 electron-proton correlation energy.

Similar to the single-component case, it is simplest to interpret the contribution of this Hugenholtz diagram by expanding it as an antisymmetrized Goldstone diagram. This expansion is not unique, and we show two possible expansions (Figures 2b and 2c). The rules for interpreting these antisymmetrized Goldstone diagrams are identical to the single-component case except that antisymmetrized one-electron one-proton integrals are used for two vertices connected by interaction lines and that resolvent lines are interpreted in terms of the electronic and protonic zeroth-order energies rather than just the electronic zeroth-order energies. The final change is that the phase factor is now $(-1)^{h-l-ep}$ where $h$ is the number of hole lines (both electronic and protonic), $l$ is the number of loops, and $ep$ is the number of interaction vertices with both electronic and protonic holes or particles lines (*i.e.*, electron-proton interactions). The dependence on the number of electron-proton interaction vertices is due to the electron-proton interaction being attractive and therefore negative rather than the repulsive electron-electron interaction as in the single-component case. In future diagrams, interaction vertices with only electronic hole and particle lines are interpreted identically to the single-component case.

Using these and the existing single-component diagrammatic rules, the antisymmetrized Goldstone diagram in Figure 2b can be written as

$$E_{\text{MP2}}^{\text{ep}} = \sum_{iIaA} \frac{|\langle iI||aA\rangle|^2}{\epsilon_i + \epsilon_I - \epsilon_a - \epsilon_A} \tag{2}$$

while the second antisymmetrized Goldstone diagram can be written as

$$E_{\text{MP2}}^{\text{ep}} = \sum_{iIaA} \frac{|\langle iI||Aa\rangle|^2}{\epsilon_i + \epsilon_I - \epsilon_a - \epsilon_A}. \tag{3}$$

To obtain the usual MP2 expression for the electron-proton correlation energy, all integrals that involve integration over a particle simultaneously in an electronic and protonic orbital are set to zero. Formally, we set equal to zero all one-electron, one-proton molecular orbitals integrals of the type

$$\langle pq|PQ\rangle = \langle PQ|pq\rangle = \langle pP|Qq\rangle = \langle Pp|qQ\rangle = 0. \tag{4}$$

Then simplifying Eqs. 2 and 3 with the definitions in Eq. 4, the usual MP2 electron-proton correlation energy is obtained

$$E_{\text{MP2}}^{\text{ep}} = \sum_{iIaA} \frac{|\langle iI|aA\rangle|^2}{\epsilon_i + \epsilon_I - \epsilon_a - \epsilon_A}. \tag{5}$$

We next derive the diagrammatic expressions for the MP3 electron-proton correlation energy. Though multicomponent MP3 has been previously derived,[8] the previous derivation was missing diagrams, so the full expression for it does not yet exist in the literature. In single-component MP3, the electron-electron correlation energy can be written diagrammatically with three directed Hugenholtz diagrams (Figure 3).

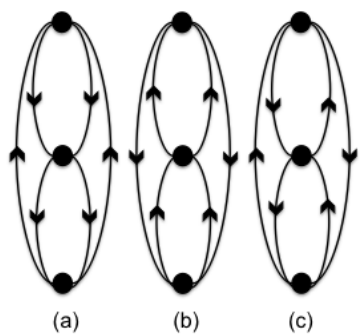

**Figure 3:** The three distinct Hugenholtz diagrams for single-component MP3.

Using the rules for deriving electron-proton Hugenholtz diagrams gives seven distinct diagrams for the MP3 electron-proton correlation energy (Figure 4).

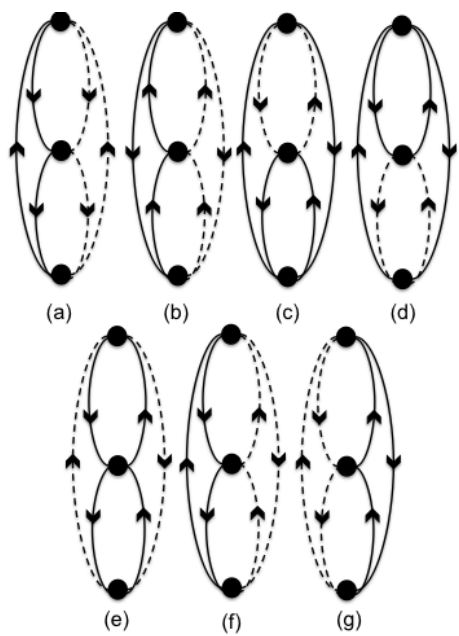

**Figure 4:** The seven distinct Hugenholtz diagrams for the electron-proton correlation energy for multicomponent MP3

The single-component diagrams (a) and (b) in Figure 3 generate one diagram each in Figure 4 ((a) and (b), respectively), while diagram (c) in Figure 3 generates the remaining five diagrams in Figure 4, which demonstrates how the number of diagrams in many-body theory for electron-proton correlation can be much greater than the number of diagrams for electron-electron correlation. We note that the previous derivation of multicomponent MP3 did not include diagrams (c) and (d) of Figure 4.

In single-component MP4, the linked form of the fourth-order electron-electron correlation energy can be written as the sum of 39 diagrammatic contributions. Applying the rules for converting these electron-electron diagrams into electron-proton diagrams gives 186 diagrams. In the interest of brevity, we do not list all these diagrams in the main text of the study but direct the reader to the complete set of diagrams in the Supplemental Material. Instead, we show one single-component triple-excitation electron-electron Hugenholtz diagram and its extension to eight electron-proton diagrams in Figure 5. This further emphasizes the large number of diagrams that result when going from a single component to a multicomponent formalism, but the diagrammatic approach is still conceptually simpler and less error prone than deriving these terms algebraically. As mentioned previously, the prior derivation of multicomponent MP4 only included fourth-order ring and ladder diagrams, which were all double-excitation contributions.

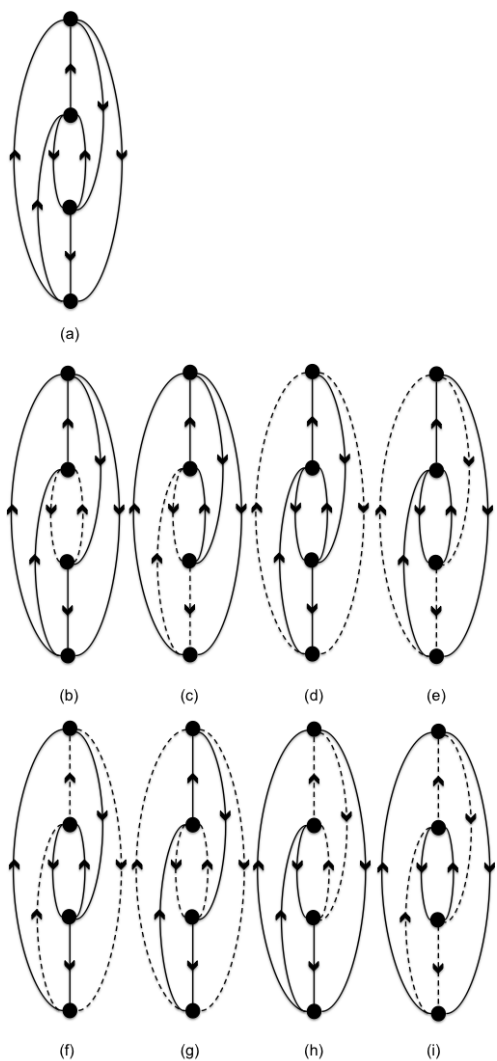

**Figure 5:** (a) A single-component fourth-order triple-excitation Hugenholtz diagram and (b)-(i) the eight electron-proton Hugenholtz diagrams derived from (a).

Like single-component MP4,[27] the most computationally expensive terms in the multicomponent MP4 equations formally scale $N^7$ with respect to system size and correspond to triple-excitation contributions to the fourth-order energy. For the electron-proton correlation energy, if we assume that the number of protonic orbitals is less than the number of electronic orbitals, which is almost always the case in multicomponent calculations, the most computationally expensive of these $N^7$ scaling terms can be further broken up into electronic and protonic contributions as $N_e^5 N_p^2$, where $N_e$ and $N_p$ are a measure of the size of the electronic and protonic subsystems, respectively. For multicomponent methods that treat only a single proton quantum mechanically, as is the case for all calculations in this study, the number of protonic MOs does not increase with system size and so the electron-proton correlation contribution to the fourth-order perturbative energy scales $N_e^5$ with respect to system size, albeit with a larger prefactor that is equal to the number of virtual protonic orbitals.

As has been previously shown in orbital-optimized MP2,[19] electron-electron and electron-proton correlation can be separated and treated using different levels of perturbation theory. A similar idea can be used for multicomponent MP4 where the electron-electron correlation is treated using MP4SDQ and the electron-proton correlation using the full MP4SDQT contribution. Though this still formally scales computationally as $N^7$ with respect to system size rather than $N^6$ as in multicomponent MP4SDQ, we expect the triples contributions to the fourth-order electron-proton correlation energy to not be a large computational expense compared to the electron-electron MP4SDQ terms due to arguments like those made in the previous paragraph. Additionally, we have implemented a method that calculates the electron-electron correlation using the full MP4SDQT contribution with the electron-proton correlation calculated at the MP4SDQ level. Though we do not expect this method to be useful in practice, it will be used to show how the inclusion of the fourth-order electron-proton triples contribution increases the accuracy of multicomponent MP4 methods.

We have implemented the necessary equations for the multicomponent MPn methods in a Python-based code that uses PySCF[38] for the calculation of the atomic-orbital (AO) integrals and the transformation of the two-particle integrals from the AO to MO basis. All equations have been spin-adapted and intermediates were defined analogous to the single-component case to increase computational efficiency.

## III. Results and Discussion

Benchmarking of the multicomponent MPn methods was performed by computing the proton affinities of a set of 12 small organic molecules with the most acidic proton treated quantum mechanically and by computing the energy of the FHF$^-$ molecule as a function of the F—F distance with the hydrogen nucleus treated quantum mechanically. Both are standard benchmarks for new multicomponent methods and have seen extensive use previously.[4] For the proton affinity benchmarking, the proton affinity of a molecule $A$ is calculated as

$$PA(A) = E_A - E_{AH^+} + \frac{5}{2}RT \tag{6}$$

where $E_A$ is the energy from a single-component calculation and $E_{AH+}$ is computed with the analogous multicomponent method.

For all proton affinity calculations, the geometries were obtained from single-component optimizations using the aug-cc-pVTZ electronic basis set[39-41] for all atoms. For the proton affinity calculations, the neecessary single-component calculations were performed with CFOUR[42] and the multicomponent calculations with our Python-based code. The multicomponent calculations used the aug-cc-pVQZ electronic basis set for the quantum proton centered at the classical position of the hydrogen atom from the single-component geometry optimization, and the aug-cc-pVTZ electronic basis set for all other atoms. The use of different electronic basis sets for different atoms is common in previous multicomponent calculations of proton affinities as electronic basis sets with high angular momentum basis functions on the quantum hydrogen atom are needed to correctly describe electron-proton correlation. Additional proton affinity calculations with the aug-cc-PVTZ or aug-cc-pVDZ electronic basis set for all atoms can be found in the Supporting Material. It can be seen from those results that without high angular momentum basis functions on the quantum hydrogen atom, the calculated proton affinities are less accurate. This is in contrast to the single-component proton affinities, which can be found in the Supporting Materials, where the electronic basis set has a much smaller effect on the mean unsigned error of the proton affinities. For the multicomponent MP3 and MP4 methods, their single-component counterparts perform better for the calculation of proton affinities, which is different than the relative performance of multicomponent and single-component CCSD. Despite this slightly worse performance, multicomponent MPn methods still offer a potential advantage over their single-component analogues as they do not require a Hessian calculation in order to include the ZPE of the quantum proton.

Additionally, the single-component MPn methods all have more similar errors in proton affinities than the multicomponent MPn methods, which indicates that much of the error in the multicomponent proton affinity calculations arises from an incomplete treatment of electron-proton correlation. All multicomponent calculations used the PB4F protonic basis set,[43] which was centered at the position of the classical hydrogen atom from the single-component geometry optimization.

Proton affinity calculations were performed using the MP2, MP3, MP4SDQ, and MP4SDQT methods for the electron-electron correlation. The electron-proton correlation was computed using the same level of theory as the electron-electron correlation except in the case of MP4SDQ and MP4SDQT, where additional calculations were performed with the electron-proton correlation calculated at the MP4SDQT and MP4SDQ levels of theory, respectively, to enable the investigation of the importance of fourth-order triple excitations in the electron-proton correlation energy.

The calculations on FHF$^-$ were performed using the same protonic basis sets as the proton affinity calculations, but with the aug-cc-pVTZ electronic basis set for all atoms. Once again multicomponent calculations were performed using the MP2, MP3, MP4SDQ, and MP4SDQT methods for the electron-electron and electron-proton correlation with additional MP4SDQ and MP4SDQT calculations performed to investigate the importance of triple excitations in the electron-proton correlation energy. For these calculations, the Fourier-grid Hamiltonian (FGH) method,[44] which is nearly numerically exact for a given one-particle electronic basis set and has been extensively used in previous multicomponent benchmarking studies, was used as a reference. The single-component calculations for the FGH calculations were performed using the CFOUR program at the single-component MP4SDQ and MP4SDQT level of theory using the aug-cc-pVTZ basis sets.

The results from the proton affinity calculations are presented in Table 1. In general, the mean unsigned error (MUE) decreases monotonically as the perturbative order of the included contributions to the correlation energy is increased. The errors of the multicomponent MP3 method are most similar to the multicomponent CCD method, while the errors of the multicomponent MP4 methods are most similar to the multicomponent CCSD method. These similarities and the relative accuracy of the methods in this study are also commonly observed in single-component quantum mechanical benchmarking. Previous benchmarking[45-46] of DFT functionals for the prediction of atomic and diatomic densities has demonstrated that single-component MP2 is a relatively inaccurate method for the prediction of electronic densities compared to other wave-function based methods and we hypothesize that the previous failures of multicomponent MP2 for the prediction of protonic densities and other properties may be more related to the well-known limitations of MP2 in a single-component framework rather than any unique challenges imposed by the failures of multicomponent HF.

Inclusion of triple excitation diagrams into the electron-proton correlation lowers the MUE when the electron-electron correlation is treated with either the MP4SDQ and MP4SDQT methods by over 45 and 35%, respectively, which demonstrates the importance of triple excitation contributions to the electron-proton correlation energy for these systems. Somewhat surprisingly, the MP4SDQ/MP4SDQT method, where the electron-electron correlation is treated at the MP4SDQ level of theory and the electron-proton correlation at the MP4SDQT level of theory, rather than the full MP4SDQT method has the lowest error of all the new methods introduced in this study. In single-component quantum chemistry, MP4SDQT is typically more accurate than both MP4SDQ and CCSD and is most similar in accuracy to CCSD(T). However, we note that the benchmark values for the proton affinities are experimental values and not full CI values and that a multicomponent CCSD(T) method has never been implemented. Therefore, no theoretical benchmarking data beyond the multicomponent CCSD level exists. It is possible that the multicomponent MP4SDQT method is better reproducing the exact theoretical or multicomponent CCSD(T) value, but due to approximations in the multicomponent scheme for calculating proton affinities,[14] the exact theoretical value disagrees with the experimental one. Further study on the topic would be a useful future research direction and with the derivation of the fourth-order triples-excitation energy in this study it should be possible to derive a (T) correction to the electron-proton correlation energy to test this possibility.

More immediately, these results demonstrate that the MP4SDQ/MP4SDQT method can calculate proton affinities as accurately as the multicomponent CCSD and provides further evidence that when protonic properties are of primary interest, the electron-electron and electron-proton correlation can be treated at different levels of theory as has been previously suggested.[19] If triple electronic excitation contributions from the electron-electron correlation energy are neglected, the order of the most computationally expensive terms with respect to the size of the electronic system is $N^6$, which is identical to multicomponent CCSD, but the MP4SDQ/MP4SDQT method is non-iterative unlike the multicomponent CCSD method. This should greatly increase computational efficiency especially because iterative multicomponent methods can take an order of magnitude more iterations to converge than a single-component calculation on an identical system.[18-19, 47] As mentioned in Section II, the overall formal scaling of the multicomponent MP4SDQ/MP4SDQT method is $N^7$, but these $N^7$ terms have at least an $N^2$ contribution from protonic-related quantities such that they have a maximum of $N^5$ contribution from electronic-related quantities. Additionally, if only a single proton is treated quantum mechanically, the $N^2$ contribution from protonic-related quantities remains fixed as the size of the system increases, which means that the $N^6$ terms in the electron-electron correlation energy will come to dominate the overall computational effort in an MP4SDQ/MP4SDQT calculation.

This behavior can be qualitatively demonstrated by looking at the relative computational effort for the multicomponent MP4SDQT calculations on the HCO and HCOOH molecules. For HCO, the fourth-order electron-proton triple-excitation contributions are 20.0% of the overall computational expense, while the fourth-order electron-electron triple-excitation contributions are 34.5%. For the HCOOH molecule, these percentages are 15.0 and 35.7%, respectively, which shows the decreasing contribution to the overall computational expense of the fourth-order electron-proton triple-excitation contributions as the system size is increased.

For the multicomponent MP4SDQ/MP4SDQT method, the fourth-order electron-proton triple-excitation contributions are 30.0 and 23.9% of the computational expense for the HCO and HCOOH molecules, respectively, which qualitatively shows that for larger molecules, the computational expense of a multicomponent MP4SDQ/MP4SDQT calculation becomes increasingly dominated by the $N^6$ terms in the electron-electron correlation energy. Such computational scaling trends should also hold true for a purely electron-proton (T) correction to multicomponent CCSD and based on the vastly improved proton affinities predictions in this study when triple-excitations are included in the electron-proton correlation energy, such an approach may be a computationally efficient way to improve the multicomponent CCSD method.

| molecule | exp[a] | HF | MP2 | MP3 | MP4SDQ | MP4SDQ/MP4SDQT | MP4SDQT/MP4SDQ | MP4SDQT | CCD[b] | CCSD[c] | CCSD (single component)[c] |
|---|---|---|---|---|---|---|---|---|---|---|---|
| $CN^-$ | 15.31 | 0.91 | 0.37 | 0.27 | 0.15 | 0.10 | 0.20 | 0.15 | 0.26 | 0.11 | 0.04 |
| $NO_2^-$ | 14.75 | 0.53 | 0.38 | 0.05 | 0.02 | 0.05 | 0.18 | 0.10 | 0.07 | 0.01 | 0.14 |
| $NH_3$ | 8.85 | 0.52 | 0.28 | 0.17 | 0.08 | 0.03 | 0.13 | 0.08 | 0.16 | 0.05 | 0.04 |
| $HCOO^-$ | 14.97 | 0.55 | 0.42 | 0.08 | 0.05 | 0.02 | 0.18 | 0.11 | 0.16 | 0.04 | 0.09 |
| $HO^-$ | 16.95 | 0.36 | 0.41 | 0.03 | 0.08 | 0.02 | 0.25 | 0.19 | 0.09 | <0.01 | 0.13 |
| $HS^-$ | 15.31 | 0.84 | 0.41 | 0.29 | 0.14 | 0.08 | 0.20 | 0.14 | 0.26 | 0.10 | 0.05 |
| $H_2O$ | 7.16 | 0.52 | 0.28 | 0.15 | 0.09 | 0.03 | 0.13 | 0.08 | 0.16 | 0.06 | 0.04 |
| $H_2S$ | 7.31 | 0.65 | 0.24 | 0.17 | 0.04 | 0.01 | 0.07 | 0.02 | 0.15 | <0.01 | 0.11 |
| $CO$ | 6.16 | 0.84 | 0.14 | 0.20 | 0.07 | 0.02 | 0.03 | 0.01 | 0.19 | 0.04 | 0.09 |
| $N_2$ | 5.12 | 0.76 | 0.28 | 0.20 | 0.09 | 0.04 | 0.10 | 0.04 | 0.19 | 0.07 | 0.09 |
| $CO_2$ | 5.60 | 0.60 | 0.30 | 0.17 | 0.07 | 0.01 | 0.14 | 0.08 | 0.14 | 0.04 | 0.07 |
| $CH_2O$ | 7.39 | 0.33 | 0.29 | 0.11 | 0.04 | 0.02 | 0.13 | 0.07 | 0.12 | <0.01 | 0.09 |
| MUE | | 0.62 | 0.32 | 0.16 | 0.08 | 0.04 | 0.14 | 0.09 | 0.16 | 0.04 | 0.08 |
| MaxAE | | 0.91 | 0.42 | 0.29 | 0.14 | 0.10 | 0.25 | 0.19 | 0.26 | 0.11 | 0.14 |

**Table 1:** Experimental proton affinities (exp), unsigned errors, overall mean unsigned error (MUE) and maximum absolute error (MaxAE) of the calculated proton affinities relative to the experimental value. All values are in eV.
[a]Refs 48-52, [b]Ref 16, [c]Ref 14

The energy of the FHF$^-$ molecule as a function of the F—F distance for the multicomponent MP4SDQ method and the MP4SDQ method with triple-excitation contributions included in the electron-proton correlation energy (MP4SDQ/MP4SDQT) is shown in Figure 6. Graphs of the energy versus the F—F distance for the multicomponent HF, MP2, and MP3 methods can be found in the Supplemental Materials. Both multicomponent methods show a similar shift in their energy curves with the minimum energy F—F distance being approximately 0.005 Å more than the reference FGH calculation with the single-component MP4SDQ method. Such behavior is similar to previous multicomponent CCSD results.[14] Inclusion of the triple-excitation contributions to the electron-proton correlation energy shifts the minimum energy F—F distance to a slightly larger value, but overall, this effect is small.

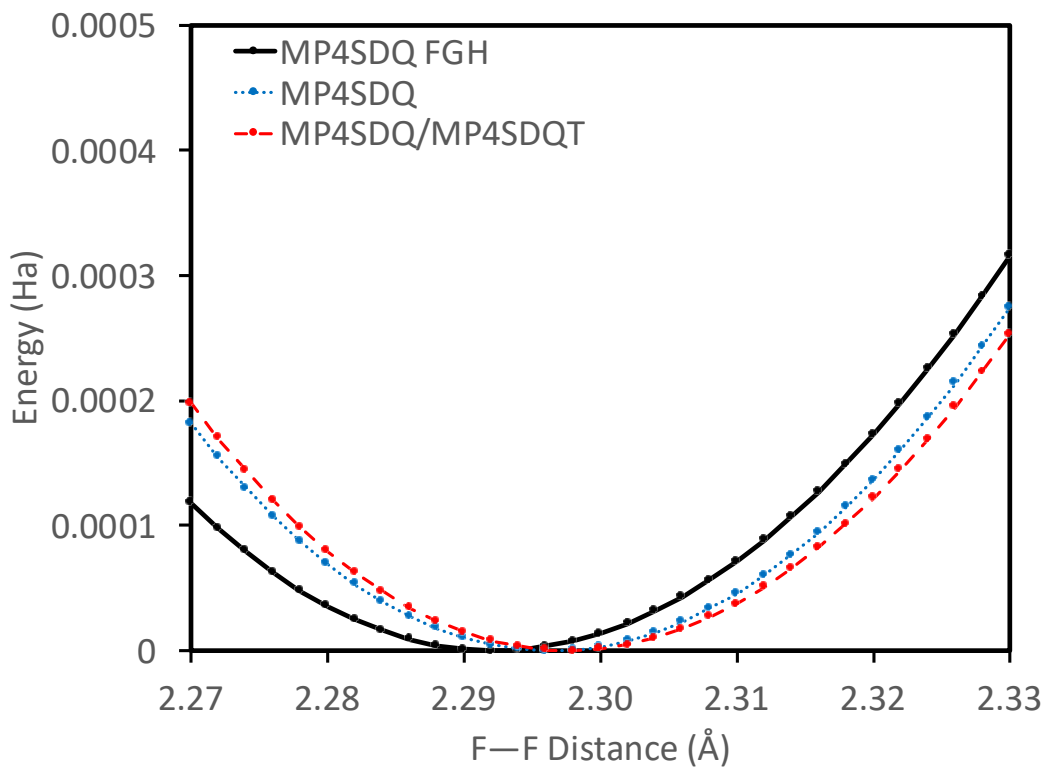

**Figure 6:** Energy of the FHF⁻ molecule as a function of the F—F distance for the multicomponent MP4SDQ (dotted blue), MP4SDQ/MP4SDQT (dashed red), and FGH method (solid black) using the single-component MP4SDQ method to compute the grid (MP4SDQ FGH). For all methods, the zero of energy is set at the minimum energy F—F distance.

The energy of the FHF$^-$ molecule as a function of the F—F distance for the multicomponent MP4SDQT method and the multicomponent MP4SDQT method with triple excitation contribution not included in the electron-proton correlation energy (MP4SDQT/MP4SDQ) is shown in Figure 7. Both the multicomponent MP4SDQT and MP4SDQT/MP4SDQ methods do a similar job of reproducing the reference MP4SDQT FGH energy curve compared to the multicomponent MP4SDQ and MP4SDQ/MP4SDQT methods. The inclusion of triple-excitation contributions to the electron-proton correlation energy again shifts the minimum energy F—F distance to a slightly larger value.

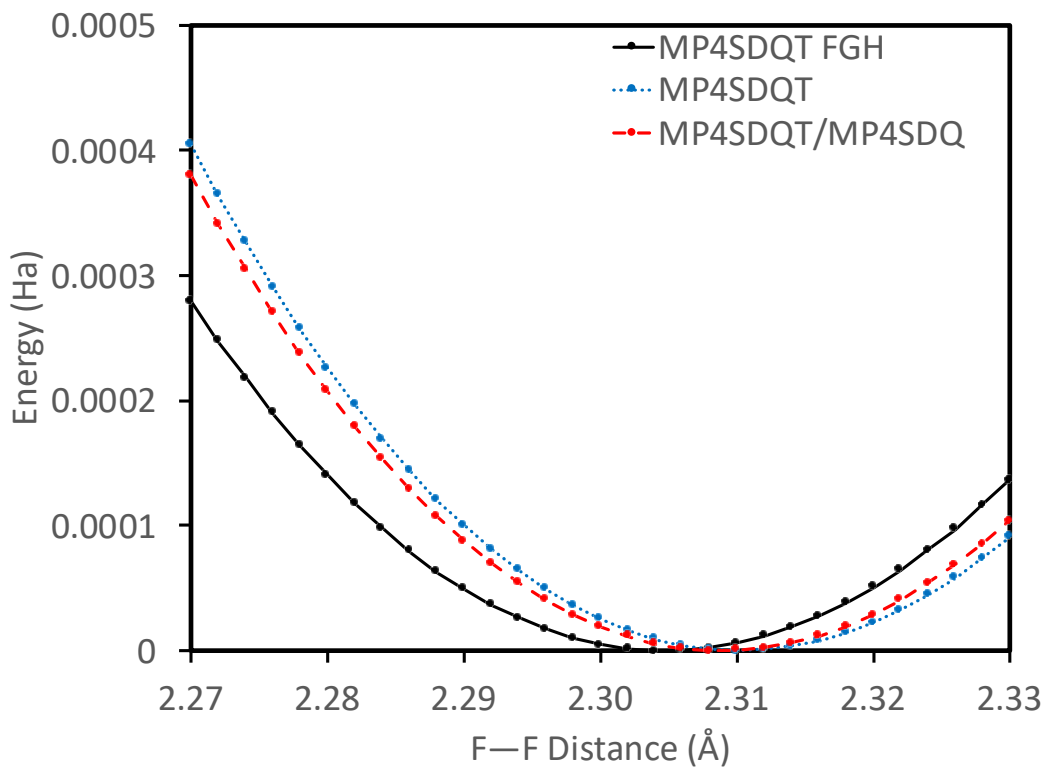

**Figure 7:** Energy of the FHF⁻ molecule as a function of the F—F distance for the multicomponent MP4SDQT (dotted blue), MP4SDQT/MP4SDQ (dashed red), and FGH method (solid black) using the single-component MP4SDQT method to compute the grid (MP4SDQT FGH). For all methods, the zero of energy is set at the minimum energy F—F distance.

**IV. Conclusions**

In this study, the full multicomponent MP3 and MP4 equations were derived and implemented for the first time. The multicomponent MP3 method was shown to perform similarly to the multicomponent CCD method while the multicomponent MP4 method was shown to perform most similarly to the multicomponent CCSD method. This performance suggests that multicomponent methods behave more similarly to their single-component counterparts than previously assumed.

The inclusion of triple-excitation contributions to the fourth-order electron-proton correlation energy is shown to significantly lower the proton affinity MUE. These fourth-order electron-proton triple contributions can be included independently of the fourth-order electron-electron contributions resulting in the MP4SDQ/MP4SDQT method. While the computational expense of the fourth-order electron-proton triple contributions formally scale as $N^7$, this $N^7$ scaling has an $N^2$ contribution from the protonic part of the system such that for a system that treats a single proton quantum mechanically, the $N^6$ purely electronic contributions are the ultimate computational bottleneck. Given the results in this study, it is suggested that for multicomponent calculations focused on computing protonic properties, a computationally efficient (T) correction to multicomponent CCSD could be derived where only triple contributions to the electron-proton correlation energy are included.

**Supplementary Material**

See supplementary material for the complete diagrammatic expressions for the multicomponent MP4 method, the graph of the energies of the FHF$^-$ molecule with the multicomponent HF, MP2, and MP3 methods as a function of F—F distance and additional single-component and multicomponent proton affinity calculations with different electronic basis sets.

## Acknowledgements

KRB thanks the University of Missouri-Columbia for Startup Funding

## Author Declarations

The authors have no conflicts to disclose.

## AIP Publishing Data Sharing Policy

The data that support the findings of this study are available within the article and its supplementary material.

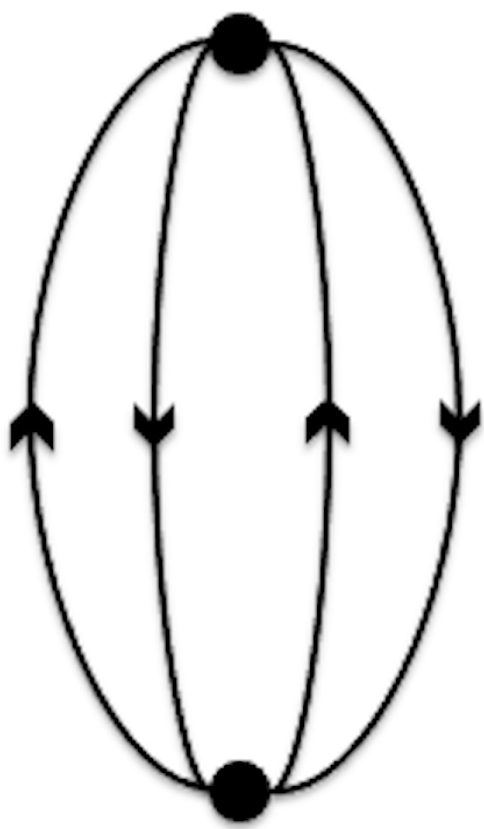 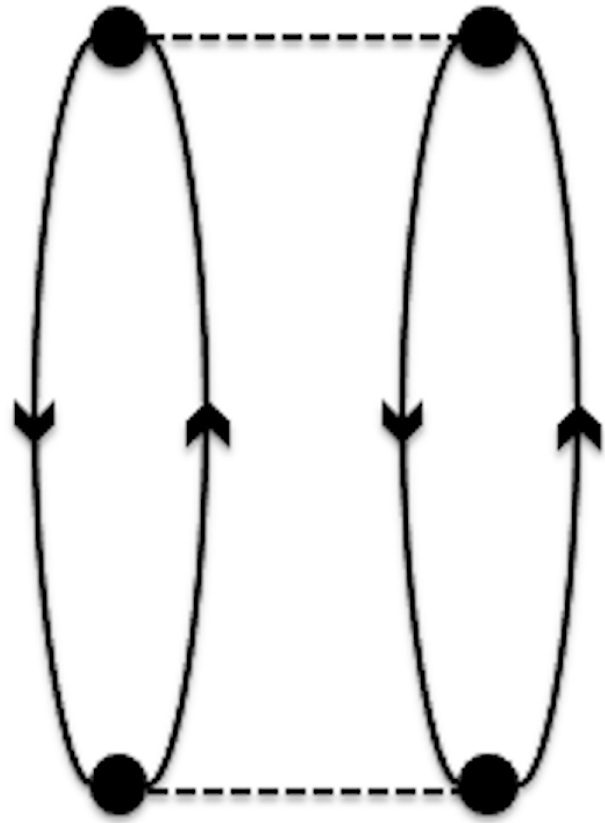

(a) (b)

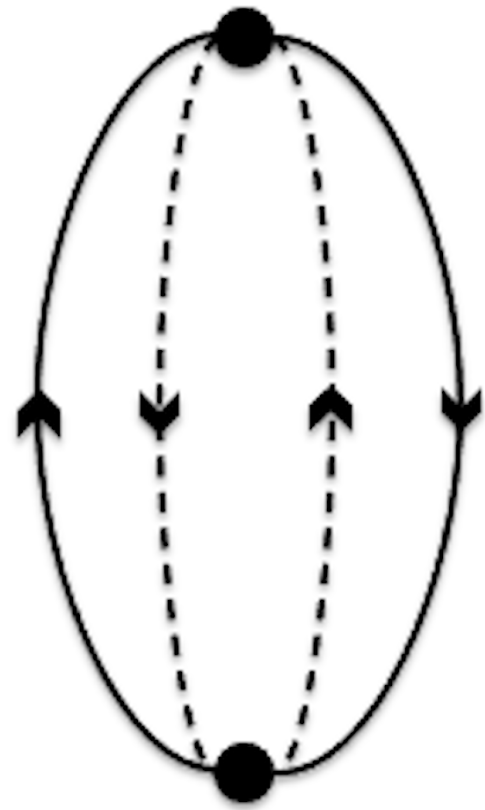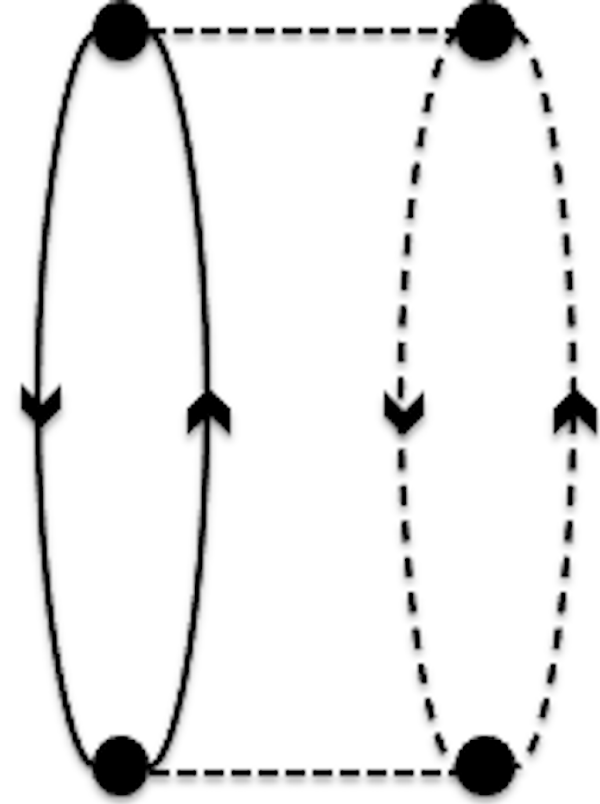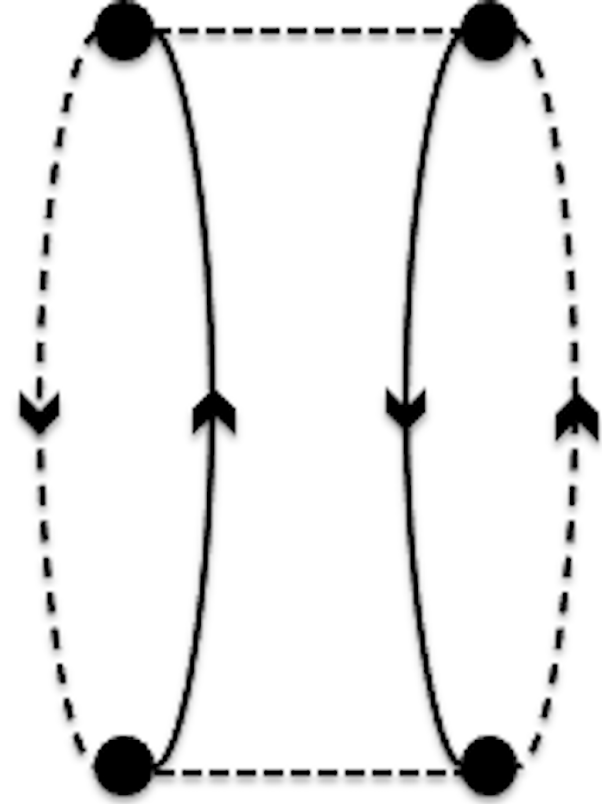

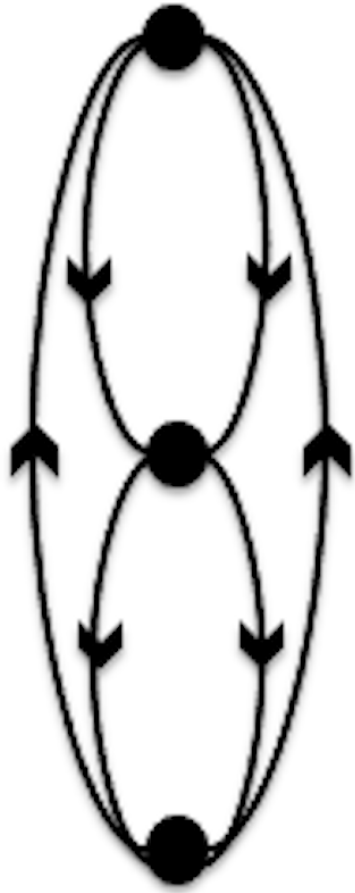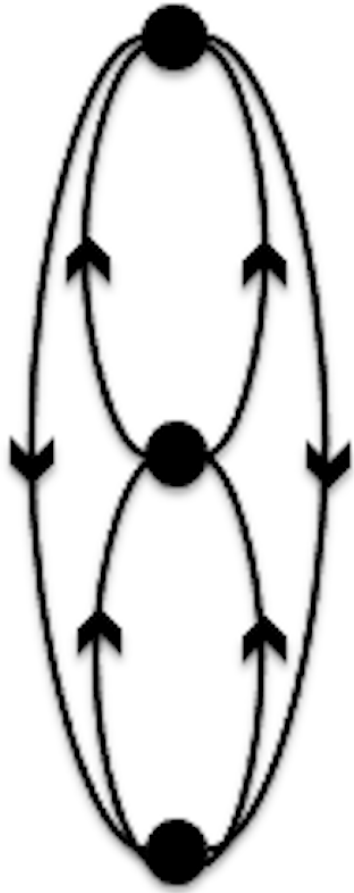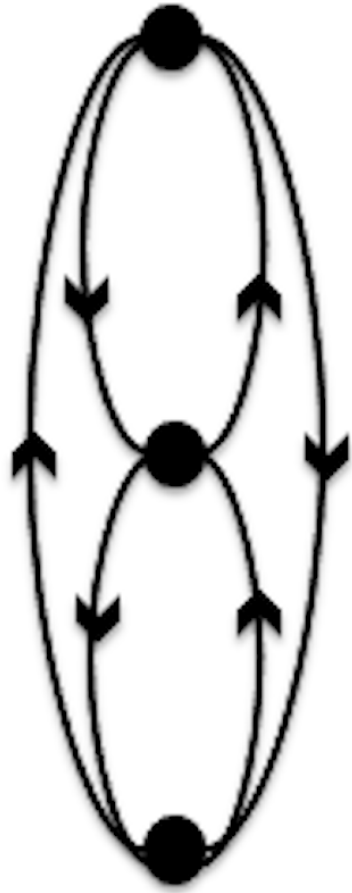

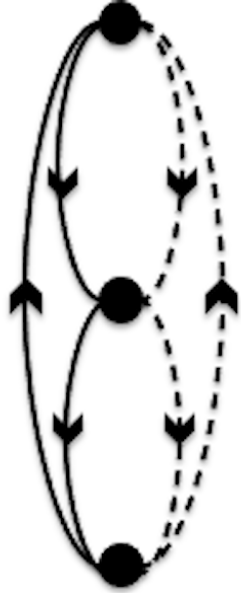 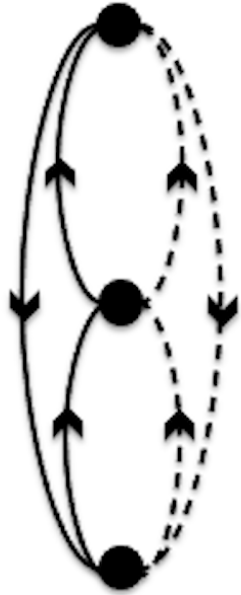 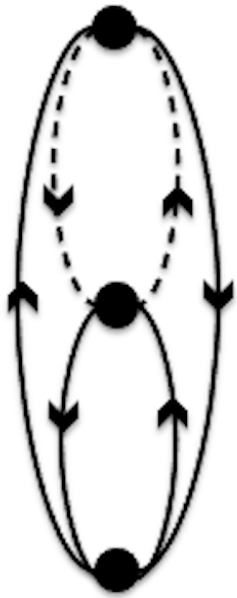 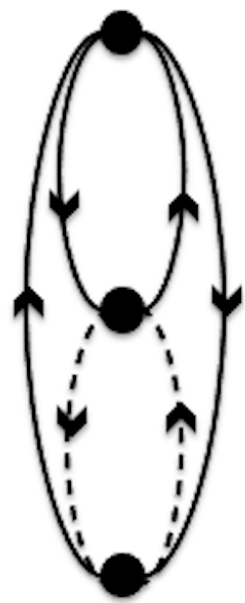

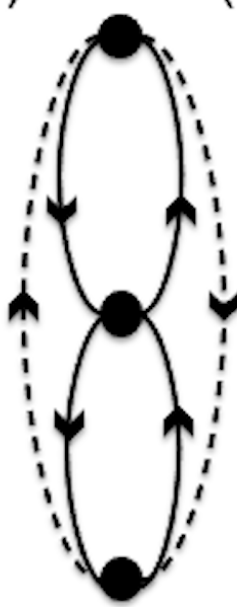 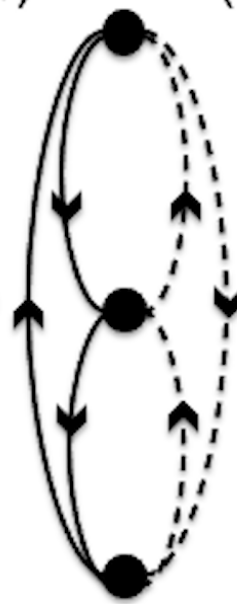 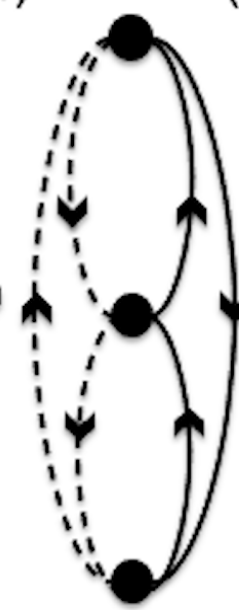

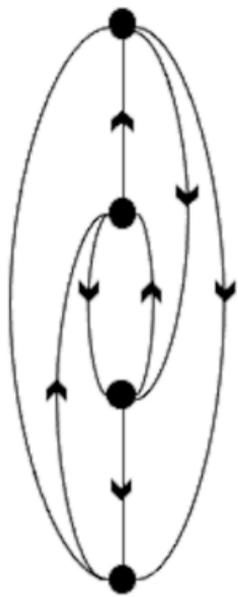

(a)

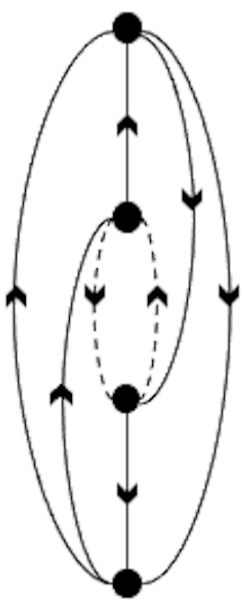 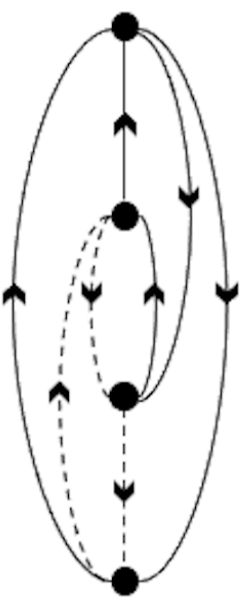 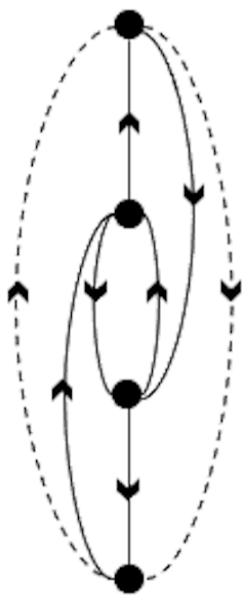 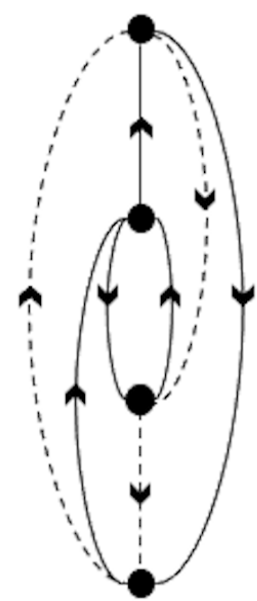

(b) (c) (d) (e)

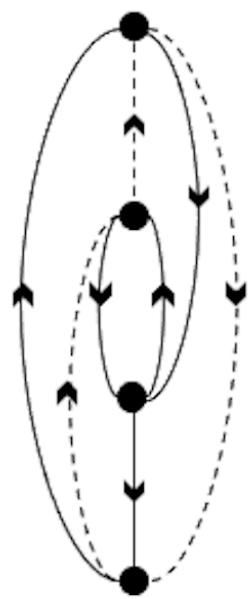 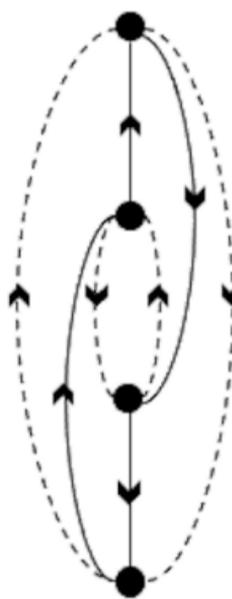 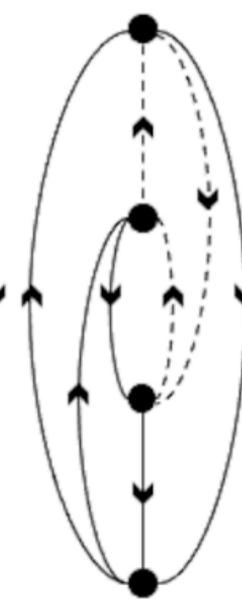 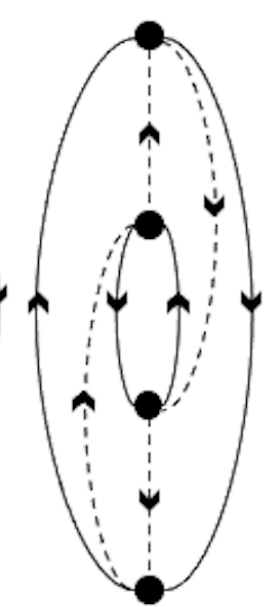

(f) (g) (h) (i)

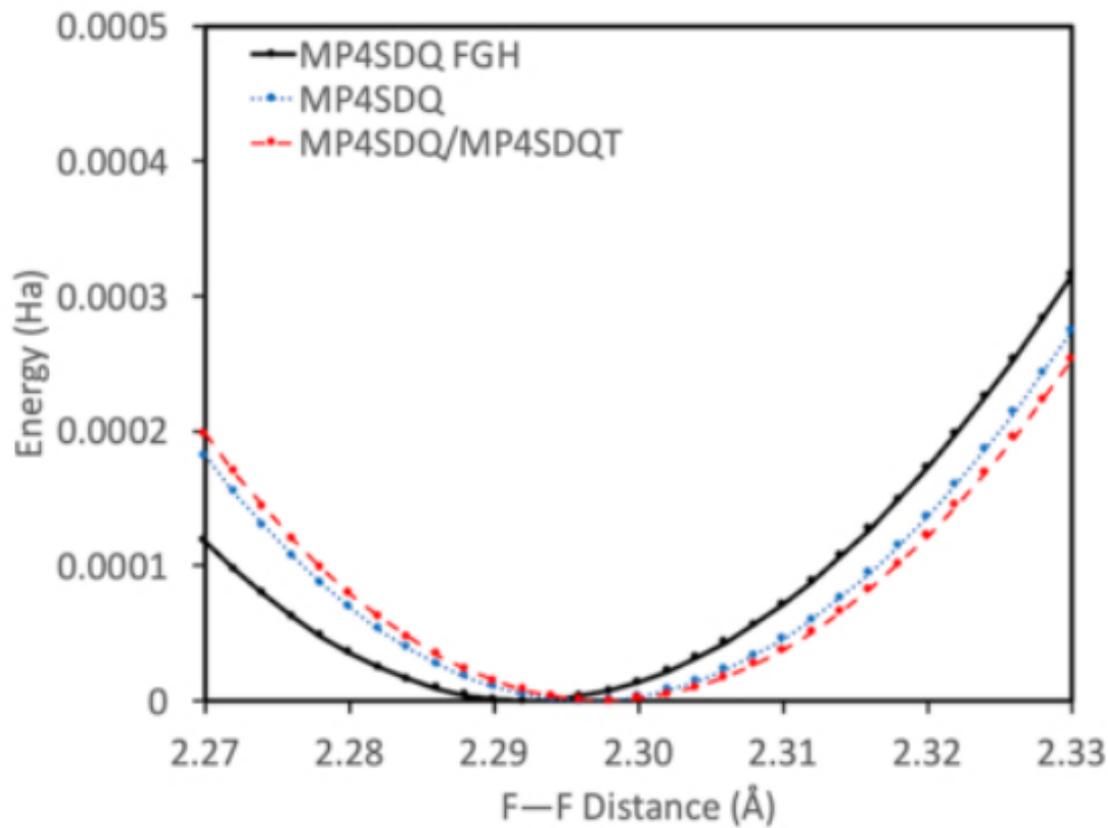

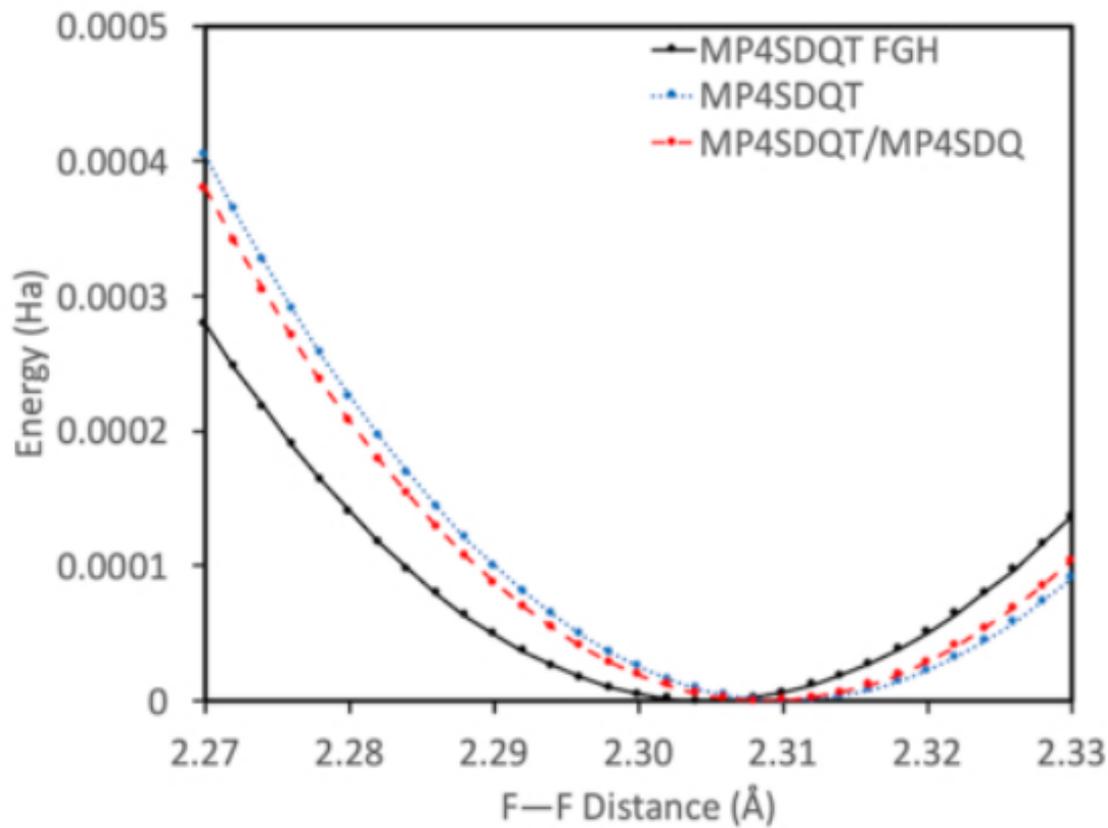